\documentclass[11pt, amsfonts]{amsart}

\usepackage{fullpage}

\usepackage{url}

\usepackage{amssymb}

\renewcommand{\a}{\alpha}
\renewcommand{\b}{\beta}
\newcommand{\g}{\gamma}
\renewcommand{\d}{\delta}
\newcommand{\D}{\Delta}

\renewcommand{\S}{\Sigma}

\newcommand{\cC}{{\mathcal C}}

\newcommand{\cT}{{\mathcal T}}

\newcommand{\cE}{{\mathcal E}}

\newcommand{\cH}{\mathcal H}
\newcommand{\cI}{\mathcal I}

\newcommand{\bR}{\mathbb R}

\newcommand{\dm}{\partial M}

\newcommand{\be}{\begin{equation}}
\newcommand{\ee}{\end{equation}}

\renewcommand{\phi}{\varphi}
\renewcommand{\epsilon}{\varepsilon}

\newcommand{\del}{\partial}

\theoremstyle{plain}

\theoremstyle{definition}

\theoremstyle{remark}

\def\blacksquare{\hbox to .60em {\vrule width .60em height .60em}}

\begin{document}

\title[ ]{On quasi-local Hamiltonians in General Relativity}

\author[ ]{Michael T. Anderson}

\address{Dept.~of Mathematics, Stony Brook University, Stony Brook, N.Y.~11794-3651, USA} 
\email{anderson@math.sunysb.edu}
\urladdr{http://www.math.sunysb.edu/$\sim$anderson}

\thanks{Partially supported by NSF grant DMS 0905159}

\begin{abstract}

We analyse the definition of quasi-local energy in GR based on a Hamiltonian analysis of the 
Einstein-Hilbert action initiated by Brown-York. The role of the constraint equations, in particular 
the Hamiltonian constraint on the timelike boundary, neglected in previous studies, is emphasized 
here. We argue that a consistent definition of quasi-local energy in GR requires, at a minimum, a 
framework based on the (currently unknown) geometric well-posedness of the initial boundary 
value problem for the Einstein equations.

\end{abstract}

\maketitle

  The analysis of the gravitational field by Arnowitt-Deser-Misner \cite{1} has led to a clear and well-defined 
construction of the Hamiltonian, and resulting definitions of energy, linear and angular momentum in the 
context of asymptotically flat spacetimes. These concepts are obviously of basic importance in understanding 
the physics of such (infinite) isolated gravitating systems. Nevertheless, infinite systems are 
idealizations of more realistic physical situations, and it is desirable to have available a similar analysis 
in the case of physical systems of finite extent. 

   However the understanding of this issue for domains of finite extent is much less satisfactory. 
Despite numerous proposals, from a number of different viewpoints, a consensus has not yet been 
reached on a suitable definition of the Hamiltonian or energy of a finite system, i.e.~a 
quasi-local Hamiltonian; cf.~\cite{2} for an excellent detailed survey of the current state 
of the art. 

   In this paper, we first examine and comment on the approach to the definition of energy of a finite 
region of spacetime based on the Hamiltonian formulation of GR. This is essentially based on a localization 
of the approach taken by ADM \cite{1} and Regge-Teitelboim \cite{3}, keeping careful track of the boundary 
terms that arise in a Hamiltonian or Hamilton-Jacobi analysis. This approach was initiated and 
pioneered by Brown-York (BY) \cite{4}. To keep the discussion focused on the central issue, 
we only consider the gravitational field, (so other matter fields are set to zero); in addition, 
we consider only the energy and not related concepts such as linear and angular momentum, 
although this could be done without undue difficulty. Finally most all of the discussion below 
applies also to more recent modifications of the BY approach by several authors, cf.~ \cite{5}-\cite{8}; 
however again for clarity and simplicity we focus on the Brown-York Hamiltonian and leave it to the 
reader to extend the analysis to the more recent alternatives. 

\medskip

  We first recall the set-up. Let $M$ be a spacetime region, topologically of the form $I \times \S$, 
with $I = [0,1]$ parametrizing time and $\S$ a compact 3-manifold with boundary $S$; typically 
$S = \del \S$ is a 2-sphere and $\S$ a 3-ball. The boundary $\dm$ of $M$ is a union of two spatial 
hypersurfaces $\S_{0} \cup \S_{1}$ and the timelike boundary $\cT = I\times \del S = \cup S_{t}$. 
These boundaries meet at the seams or corners $S_{0}$ and $S_{1}$. The Einstein-Hilbert action 
is then given by (setting $8\pi G = 1$),  
\be \label{1}
\cI_{EH}({\sf g}) = \int_{M}R_{\sf g}dV_{\sf g},
\ee
where $\sf g$ is a smooth Lorentz metric on $M$.  

  The Hamiltonian in GR plays two important but a priori distinct roles. In classical field theories 
without dynamical gravity and based on a fixed background (Minkowski) spacetime, these two 
roles coincide. It is not at all clear, at least in the case of finite domains, whether they can be 
made to coincide in GR. 

  A Hamiltonian $\cH$ for the action \eqref{1} depends on a choice of time function $t$ and 
associated vector field $\del_{t}$, giving a foliation $\S_{t}$ of the spacetime, i.e.~a $3+1$ 
decomposition. Given the spacetime metric ${\sf g}$, this is equivalent to specifying a lapse $u$ 
and shift $X$, so that $\del_{t} = uT + X$, where $T$ is the unit timelike normal to the foliation. 
Thus $\cH = \cH_{(u, X)}$. 

  A full-fledged Hamiltonian analysis requires a well-defined phase space $T^{*}Q$, the 
cotangent bundle of the configuration space $Q$, with variables $(g, \pi)$ where $\pi$ is 
the momentum conjugate to $g$; $g$ is a Riemannian metric on $\S$. In the case of finite 
boundaries, boundary conditions for the variables $(g, \pi)$ must be specified in such a way 
that the Hamiltonian $\cH_{(u, X)}: T^{*}Q \to \bR$ is first a smooth function on $T^{*}Q$, 
and second the Hamiltonian vector field on $T^{*}Q$ (generated from $\cH$ by the 
symplectic structure on $T^{*}Q$) generates exactly the equations of motion; integral 
curves of the Hamiltonian vector field give vacuum solutions of the Einstein equations 
satisfying the boundary conditions. For this to be consistent, boundary conditions 
for the lapse-shift $(u, X)$ must be determined, and these must also be preserved 
under the equations of motion. If the above can be accomplished, one obtains a Hamiltonian 
depending on the gauge choice $(u, X)$. 

  On the other hand, in other classical field theories, the energy is understood as the Noether 
charge associated to time-translation symmetries; for Minkowski backgrounds, the energy is 
thus the time component of an invariantly defined energy-momentum 4-vector. In GR, one 
has typically no symmetries, i.e.~Killing fields, and so no conserved charges. Nevertheless, 
one can attempt to define preferred or distinguished ``quasi-symmetries" or ``quasi-Killing fields". 
This gives a preferred choice of the lapse-shift $(u, X)$, and the Hamiltonian is then taken with 
respect to such a choice, cf.~\cite{2}, \cite{9} for further discussion. 

\medskip

  The approach of Brown-York is to choose Dirichlet boundary conditions for the metric $\g = \g_{\cT}$ 
induced on $\cT$. This choice naturally conforms to the modification of the EH action by the addition 
of a boundary term - the well-known Gibbons-Hawking-York or $tr K$ boundary term \cite{10}, \cite{11}. 
Thus consider the modified Lagrangian
\be \label{2}
\cI(g) = \int_{M}R_{\sf g}dV_{\sf g} + 2\int_{\cT}kdV_{\g},
\ee
where $k = tr K$ is the mean curvature of the boundary $\cT$ in $(M, {\sf g})$ with respect to the 
outward unit normal and $\g$ is the metric on $\cT$ induced by ${\sf g}$. A straightforward 
calculation shows that the variation of $\cI$ at ${\sf g}$ in the direction $h$ is given by
\be \label{3}
\d_{\sf g}\cI (h) = -\int_{M}\langle E, h \rangle dV_{\sf g} - \int_{\cT}\langle \pi_{\cT}, 
h^{T}\rangle dV_{\g},
\ee
where $E({\sf g}) = Ric_{\sf g} - \frac{R}{2}{\sf g}$ is the Einstein tensor, $\pi_{\cT} = K - k\g$ 
is momentum conjugate to $\g$ on $\cT$ and $h^{T}$ is the induced variation of $\g$ on $\cT$; 
$\langle \cdot , \rangle$ is the pairing induced by the metric ${\sf g}$.  Thus, the variation 
vanishes on-shell, when the metric $\g$ is held fixed at $\cT$, ($h^{T} = 0$). The action is 
functionally differentiable (in fact $C^{\infty}$ smooth) with respect to variations of the metric 
fixed on the boundary. In other words, one has a well-defined variational principle for Dirichlet boundary 
data. (The variational formula \eqref{3} should also include other terms at $\S_{0}$ and $\S_{1}$ 
and the corners, but these will be ignored here since they play no role in the analysis). 

   On-shell, i.e.~on the space of solutions of the vacuum Einstein equations, the BY quasi-local Hamiltonian 
(or quasi-local energy) is formed by taking the on-shell variation of the action \eqref{2} in the direction 
of a time-like unit normal vector field $T$ on the boundary $S = \del \S$; one assumes here that 
$T$ is tangent to $\cT$.  Since $\pi(T,T) = H$, where $H$ is the mean curvature of $S \subset \S$, 
the on-shell variation in the $(T,T)$ direction is given by
\be \label{4}
-\int_{S}HdV_{\g_{S}},
\ee
where $\g_{S}$ is the metric induced on $S$. More generally, let $(u, X)$ be the lapse-shift of the 
foliation $\S_{t}$. A standard Hamiltonian analysis gives 
\be \label{5}
\cH_{(u,X)} = -\int_{S}[uH - \pi(X,\nu)]dV_{\g_{S}},
\ee
on-shell, where $\nu$ is the outward unit normal of $S \subset \S$. The case \eqref{4} is recovered 
by setting $(u, X) = (1,0)$. There remains a freedom in specifying the zero-point energy; the 
prescription of Brown-York is to define
\be \label{6}
\cH_{BY}(S) = \int_{S}(H_{0} - H)dV_{\g_{S}},
\ee
where $H_{0}$ is the mean curvature of an isometric embedding of $(S, \g_{S})$ into Euclidean 3-space 
$\bR^{3}$.  Hence $\cH_{BY}(S)$ is well-defined only if there exists a unique isometric embedding 
into $\bR^{3}$; this is the case if for instance the Gauss curvature $K_{\g_{S}}$ is positive, by the 
Weyl embedding theorem. 

\medskip

  The BY quasi-local Hamiltonian has a number of important and interesting properties, both physically and 
mathematically. The expression \eqref{4} is local on $S$ and easily computable, although the subtraction 
term in \eqref{6} is more complicated since it depends on the global structure of $(S, \g_{S})$. Particularly 
noteworthy is the result of Shi-Tam \cite{12} that for time-symmetric data ($\pi_{\S} = 0$), if the Gauss 
curvature $K_{\g_{S}} > 0$ then $\cH_{BY}(S) \geq 0$ with equality if and only if $\S$ is flat. However, 
for general surfaces $S$, $\cH_{BY}(S)$ may be negative. 

    Observe that $\cH_{BY}(S)$ depends only on the Cauchy data on the initial surface $\S$; in fact it 
depends only on the metric $g$ on $\S$ near $S$. It depends on the choice of $\S$, (within the 
domain of dependence) and one obtains different energies for different Cauchy surfaces, i.e.~the 
Brown-York energy is gauge dependent. However, it does not in fact depend on the actual unit normal 
$T$ at $S$; this is due to a cancelation of boundary terms. Note that the Cauchy data $(g, \pi)$ of $\S$ 
do not determine the unit normal $T$ at $S$; equivalently the lapse-shift $(u, X)$ at $S$ are undetermined 
by Cauchy data. 

   It follows that given an initial data surface $\S$, the BY Hamiltonian is independent of the structure of 
the spacetime outside the domain of dependence $D(\S)$ of $\S$; it is the same no matter what the metric 
is outside $D(\S)$. It is not a priori clear why the energy of the gravitational field should be independent 
of its structure at $\cT$. 

\medskip

   Although certainly natural, there is a basic problem with a definition of the Hamiltonian depending on 
the choice of Dirichlet boundary data as in and following \eqref{2}.  As stated explicitly in \cite{4}, the action 
evaluated on a classical solution (i.e. the on-shell action) is understood to be a function of the boundary metric 
$\g$ on $\cT$.  The space $\cE$ of solutions of the Einstein equations is thus to be smoothly parametrized by 
the space of metrics  $Met(\cT)$ on the boundary $\cT$, together with the space $\cC$ of Cauchy data on 
$\S$ (satisfying the constraint equations): 
$$\cE \simeq \cC \times Met(\cT).$$
However, this is not the case. In fact for generic choices of boundary metric $\g$, there will be {\it no} solution 
of the equations of motion inducing $\g$ on $\cT$. This is due the constraint equations, and in particular to 
the Hamiltonian or scalar constraint, on $\cT$. This issue, that there may be no solutions to the equations of 
motion on the phase space without the correct boundary conditions, is exactly the underlying issue and theme 
in the Regge-Teitelboim analysis of the ADM Hamiltonian. 

\medskip

  To illustrate this point clearly, we discuss it in three different but related situations. 
    
(I).  Consider first the simpler case of the pure Cauchy problem for the Einstein equations. Here the 
Cauchy data $(g, \pi)$ on $\S$ parametrize the space of solutions, and one may ask if one may specify the 
metric $g$ (Dirichlet data) arbitrarily on $\S$ to generate a solution. However, this is not the case 
since the data must satisfy the momentum constraint $\d \pi = 0$ and, more importantly here, the 
Hamiltonian or Gauss constraint $|\pi|^{2} - \frac{1}{2}(tr \pi)^{2} - R_{g} = 0$. The initial data 
$(g, \pi)$ are usually assumed to lie in Sobolev spaces $H^{s}\times H^{s-1}$, (or some analogue). 
The Hamiltonian constraint then implies $R_{g} \in H^{s-1}$.  For generic $g \in H^{s}$, $R_{g} 
\in H^{s-2}$, not in $H^{s-1}$. Hence $g \in H^{s}$ cannot be freely prescribed. Thus, as in the 
Lichnerowicz approach to solving the constraint equations, only for example the conformal class 
$[g]$ can be prescribed. 

(II). Consider next the Euclidean situation, i.e. the Euclidean Einstein equations on a compact 
domain $M$ with boundary $\dm$. Here the Einstein equations $E(g) = 0$ with Dirichlet 
boundary conditions $\g = g|_{\dm}$ given, do {\it not} form a well-posed elliptic boundary 
value problem (for any choice of gauge) essentially for the same reasons described in (I), 
namely the Gauss or Hamiltonian constraint. This is in spite of the fact that the variational 
problem for the action \eqref{2} is well-defined for DIrichlet boundary data, exactly as in 
the Lorentzian case. 

   There are many choices of boundary data which can be used to obtain a well-posed elliptic 
boundary value problem; geometrically perhaps the most natural is that given by
\be \label{7}
([\g], k),
\ee
i.e.~prescribing the conformal class $[\g]$ of the metric on $\dm$ and the mean curvature 
$k$ of $\dm$ in $M$; cf.~\cite{13} for proof of these results and further discussion. 

 (III).  The same argument applies for a time-like boundary $\cT$. The ``Hamiltonian 
constraint'' $|K|^{2} - k^{2} + R_{\g} = 0$ along $\cT$ again constrains the freedom of 
the metric $\g$ on $\cT$. For example, given a 3+1 decomposition of the spacetime, 
let $(g_{t}, \pi_{t})$ be the curve of Cauchy data on the surfaces $\S_{t}$. Working again 
in Sobolev spaces $H^{s}$, the trace of the metric (or any function) on the boundary, loses 
half a derivative, so on $\cT$, $\g_{S} \in H^{s-1/2}$. The second fundamental form $K$ of 
$\cT$ involves a (spatial) derivative of $g$, so $K \in H^{s-3/2}$. The constraint 
equation then gives $R_{\g} \in H^{s-3/2}$. But for generic $\g \in H^{s-1/2}$, one 
will not have $R_{\g} \in H^{s-3/2}$ but instead $R_{\g} \in H^{s-5/2}$ - the same 
behavior as in (I) or (II). 

  As noted above, the quasi-local energy $\cH_{BY}(S)$ is the value of the on-shell Hamiltonian 
that generates unit time translation orthogonal to $S$ at the boundary $\S$ and is given by the 
variation of the $Tr K$ action in a unit timelike normal direction along the boundary.  From 
the point of view of the initial boundary value problem (IBVP) for the vacuum Einstein 
equations, this presupposes that there exists a solution of the Einstein equations in 
the bulk $M$, whose boundary metric is of the form
\be \label{8}
\g_{\cT} = -dt^{2} + \g_{t},
\ee
on $\cT$, at least to 1st order in $t$, so that one is prescribing the form of the 
metric, i.e.~Dirichlet boundary data, on $\cT$. The Cauchy data $(g, \pi)$ at $t = 0$ 
determine the derivative $\del_{t}\g_{\cT}|_{t=0}$. However, the boundary value 
problem \eqref{8} cannot be solved in general; the IBVP is not well-posed (even at the 
linearised level) for Dirichlet boundary data. The Hamiltonian constraint serves as an 
obstruction to solvability in general. 

   In sum, the constraint equations on the Cauchy surface $\S$ and boundary $\cT$ 
"generate" the diffeomorphisms and impose constraints on the allowed Cauchy and boundary data. 
The Hamiltonian constraint on $\cT$ generates diffeomorphisms normal to $\cT$ and so is 
related to the location of $\cT$ in the spacetime which thus cannot be fully prescribed a priori. 
The presence of the constraints on $\cT$ and the difficulties they present in obtaining a 
well-behaved quasi-local Hamiltonian has not been addressed previously in the literature. 

   For the flow of the Hamiltonian vector field to be well-defined, it is necessary that the IBVP 
for the Einstein equations is well-posed. It is this criterion one should choose to seek 
suitable boundary conditions on $\cT$. 

\medskip

  Before discussing the problem in general, we consider some constructions of possible 
alternatives to the BY Hamiltonian. First, the linearisation of the scalar curvature $R_{\sf g}$ in 
the direction of a variation $h$ of ${\sf g}$ is given by $L(h) = R'_{h} = -\Box tr h + \d \d h - 
\langle Ric, h \rangle$, where $\d h$ is the divergence of $h$. Hence
\be \label{9}
\d \cI_{EH}(h) = \int_{M} -\Box tr h + \d \d h - \langle Ric, h \rangle + 
{\tfrac{1}{2}}R tr h = -\int_{M}\langle E, h \rangle + \int_{\dm}-\nu(tr h) - (\d h)(\nu), 
\ee
where $\nu$ is the unit outward normal to $\dm$ in $M$. One has $-\nu(tr h) - (\d h)(\nu) = 
-2k'_{h} - \langle K, h \rangle + \d(h(\nu))^{T})$. The last term is a divergence term, which 
integrates to 0 on the boundary, so
$$\d \cI_{EH}(h) = -\int_{M}\langle E, h \rangle - \int_{\dm}2k'_{h} + \langle K, h \rangle.$$
Since $(kdV_{g})'_{h} = k'_{h}dV_{g} + \frac{1}{2}k\, tr h dV_{g}$, this gives the variational 
formula for $\cI$ in \eqref{3}. 

  Consider next for instance the variational problem for fixed conformal class and mean curvature, 
as in \eqref{7} above. One has $\langle K, h \rangle = \langle K_{0}, h_{0} \rangle + \frac{1}{3}k\, tr h$
where $K_{0}$ and $h_{0}$ denote the trace-free parts. On other hand, $\frac{2}{3}(kdV)'_{h} = 
\frac{2}{3}k'_{h} + \frac{1}{3}k\, tr h$, and so for the action
$$\cI_{Ck}(g) = \int_{M}R + {\tfrac{2}{3}}\int_{\cT}k,$$
one has
\be \label{10}
\d \cI_{Ck}(h) = -\int_{M}\langle E, h \rangle  - \int_{\cT}[\langle \pi_{0}, h_{0}\rangle + 
{\tfrac{4}{3}}k'_{h}] .
\ee
This gives a well-defined variational problem with prescribed conformal class $[\g]$ and mean curvature $k$ 
on $\cT$, i.e.~this action is a smooth function on the configuration space of 4-d metrics with boundary 
data \eqref{7} fixed.  

  In the case of Euclidean signature, the boundary data $([\g], k)$ form a well-posed elliptic boundary value 
problem, so that one may expect to find a unique solution, (at least under mild conditions). Whether this holds for 
Lorentzian signature is unknown, (but unlikely). Suppose nevertheless for the sake of argument that the Lorentzian 
problem is well-posed. If one takes a timelike vector field $\del_{t}$ to form a Hamiltonian then linearised 
boundary data $(h_{0}, k'_{h})$ on $\cT$ determine a unique bulk $h$ solution along $\S$ (or $\S_{t}$); from this 
one may then read off Dirichlet boundary data $h^{T}$ on $\cT$. To evaluate the Hamiltonian, one then 
chooses $h$ (if possible) so that $h^{T} = T\cdot T$ at $S$. The corresponding data $(h_{0}, k'_{h})$ are then paired 
with the coefficients $(\pi_{0}, \frac{4}{3})$ above, allowing one to determine the corresponding quasi-local 
Hamiltonian. The determination of the Hamiltonian is thus ''global" on the solution and is rather complicated.

\medskip

  The simplest solutions of the Einstein equations are the time-independent static solutions, of the form
$${\sf g} = -u^{2}dt^{2} + g,$$
where $\del_{t}$ is a (hypersurface orthogonal) Killing field. Consider then a Hamiltonian analysis for 
static metrics. In place of the EH action \eqref{1} or the BY action \eqref{2},  consider here
$$\cI_{St}(g) = \int_{M}R_{\sf g}dV_{\sf g} + 2\int_{\dm}\nu(u)dV_{\g},$$
where $\nu$ is the outward unit normal. The term $R_{g}$ may be computed in terms of $R_{g}$ 
and $u$, and via an integration by parts it is easily verified that
\be \label{11}
\cI_{St}(g) = \int_{\S}uR_{g}dV_{g},
\ee
so that the boundary term disappears on passage to the Cauchy surface. A straightforward computation along the 
lines of \eqref{9}-\eqref{10} above gives
\be \label{12}
\d_{g} \cI_{St}(h, u') = \int_{\S}[\langle L^{*}u + {\tfrac{1}{2}}uR\g, h \rangle + Ru'] + \int_{S}\langle uK - \nu(u)\g_{S}, 
h^{T}\rangle + 2uk'_{h},
\ee
cf.~\cite{14} for a proof. Here $L^{*}u = D^{2}u - \D u \cdot \g - uRic$ is the adjoint of the linearisation $L$ of the 
scalar curvature. The vanishing of the bulk term in \eqref{12} gives exactly the static vacuum Einstein 
equations, as expected.  The boundary term vanishes when $h^{T} = 0$ and $k'_{h} = 0$. The mean curvature 
$k$ of $\cT$ in $M$ is the same as the mean curvature $H$ of $S$ in $\S$. 

   This gives a well-defined variational problem for the boundary data 
\be \label{13}
(\g_{S}, H) \ \ {\rm at} \ \  S = \del \S,
\ee
so that the Lagrangian is a smooth function on the configuration space with these boundary data. Note that 
one may (trivially) Wick rotate static spacetimes to Euclidean signature. In contrast to Dirichlet 
boundary data $(\g_{S}, u)$, the boundary data \eqref{13} are well-posed, i.e.~elliptic, for the static 
Einstein equations, (cf.~\cite{15}). 

    The time function $t$ and vector field $\del_{t}$ give the natural lapse-shift $(u, 0)$ for static metrics. 
With respect to this, the static Hamiltonian $\cH_{St}$, given by the variation of the action 
in the direction of the unit normal, is just given by \eqref{11}. In more detail, for $h = T\cdot T$ on 
$\cT$, one has $h^{T} = 0$ and $k'_{h} = 0$, and so, on-shell, one has
$$\cH_{St} = 0.$$

  Thus, in this situation, the quasi-local energy is zero. Physically, the simplest and most natural energy of a 
static vacuum system where one has a preferred timelike Killing field is given by the Komar energy or 
mass; it is clear that the Komar mass vanishes for compact bodies $S = \del \S$ as above. Thus, the 
energy $\cH_{St}$ agrees with the Komar energy. However, by the result of Shi-Tam \cite{12}, the BY 
energy of a static vacuum solution is strictly positive (when $K_{\g_{S}} > 0$) unless the solution is flat.

\medskip

  Turning now to the general problem one would like to find a boundary term $B(g, K)$ such that the 
action
\be \label{14}
\int_{M}R + \int_{\cT}B(\g, K)
\ee
gives a well-defined variational problem, for a choice of boundary data on $\cT$. Next, one would 
like to find an associated Hamiltonian 
\be \label{15}
\cH_{(u, X)} = \int_{\S}uC + X^{\mu}C_{\mu} + \int_{S}B_{(u,X)}(g, \pi),
\ee
where $C$ and $C_{\mu}$ are the Hamiltonian and momentum constraints respectively. Then 
boundary conditions are specified for the variables $(g, \pi)$ on the phase space $T^{*}Q$ such that 
the Hamiltonian is functionally differentiable (smooth as a function on $T^{*}Q$) and  the Hamiltonian 
vector field generates the equations of motion $E = 0$, so that one has a well-posed IBVP (in some gauge). 
Finally, if possible, one would like to select a preferred choice of time-like vector field $\del_{t}$ 
(i.e.~a preferred lapse-shift) giving a choice of quasi-Killing field. 

\medskip

     Given a choice of lapse-shift $(u, X)$ the EH action \eqref{1} decomposes into a time-space 
integral when the scalar curvature $R$ is expressed in terms of the data $(g, \pi)$ on the phase 
space $T^{*}Q$; the spatial integral is then just the integrated constraint operator given by
$$C(u, X) = \int_{\S}u(R - \pi^{2} + {\tfrac{1}{2}}(tr \pi)^2) - \langle \d \pi, X \rangle.$$
Integrating the last term by parts gives
$$C(u, X) = \int_{\S}[u(R -\pi^{2}+ {\tfrac{1}{2}}(tr \pi)^2) -\langle \pi, \d^{*}X \rangle ]
- \int_{S}\langle \pi(X), \nu \rangle.$$

  Now consider the variation of $C$ on the phase space, so with respect to $(g, \pi)$. Calculating 
the variation of $R$ and performing an integration by parts gives rise to the usual Einstein 
evolution equations in Hamiltonian form in the bulk, together with a boundary term equal to
\be \label{16}
\int_{S} u\langle \nabla_{\nu}h, g \rangle + u\langle \nu, \d h \rangle + \langle h, du\cdot \nu \rangle + 
\langle h, 2X\cdot \pi(\nu)\rangle - \langle h, \pi \rangle \langle \nu, X \rangle + 2\langle \pi', X\cdot \nu \rangle,
\ee
cf.~\cite{9} for example. Observe that the first three terms involve only the lapse $u$ while the last three terms 
involve only the shift $X$. A well-defined variational problem then holds for \eqref{14} or \eqref{15} with 
$B = 0$ provided the boundary term \eqref{16} vanishes. In general, sum of the variation of $B$ in \eqref{15} 
and \eqref{16} must vanish. 

   The first three terms in \eqref{16} can be expressed in terms of the induced metric $\g_{S}$ on $S$ and the 
mean curvature $H$ of $S \subset \S$.  In fact, (up to signs) the first three terms can be rewritten as
$$\int_{S}\langle uK - \nu(u)\g_{S}, h^{T}\rangle + 2uk'_{h},$$
exactly as the boundary term in \eqref{12}. Consider then as configuration space the space of metrics 
with shift $X = 0$, i.e.~the space of metrics of the form
$${\sf g} = -u^{2}dt^{2} + g_{t},$$
with $u = u(t, x)$. Imposing the boundary conditions \eqref{13} then gives a well-defined variational problem 
for \eqref{14} with $B = 0$, again with zero Hamiltonian on-shell. 

   This example, as well as the examples discussed previously, give quasi-local Hamiltonians coming from a 
well-defined variational principle with corresponding boundary conditions. However, in each case the Hamiltonian 
vector field is not integrable, i.e.~the associated flow equation is not generally solvable.

\medskip

  If the Hamiltonian vector field $(\frac{\del H}{\del \pi}, -\frac{\del H}{\del \g})$ is to have a well-defined flow on 
the phase space $T^{*}Q$, then there must exist a gauge choice (and in particular a choice of lapse-shift $(u, X)$) 
such that the IBVP for the Hamiltonian evolution equations are well-posed, i.e.~one has existence and uniqueness 
of solutions with given Cauchy and boundary data, and smooth dependence of the solutions on such data. 
Further, the IBVP must be geometric in the sense that solutions are isometric if and only if the Cauchy and 
boundary data differ by the action of diffeomorphisms. 

   However, as clearly stressed by Friedrich \cite{16}, it is currently a basic open problem if in fact there exists 
a choice of gauge and boundary data such that the IBVP is geometrically well-posed in this sense. There is 
a well-posed formulation of the IBVP first discovered by Friedrich-Nagy \cite{17} and a more recent formulation 
due to Kreiss-Winicour \cite{18}, \cite{19}. The results in \cite{18}, \cite{19} in particular are naturally 
formulated in harmonic gauge for the spacetime metric ${\sf g}_{\a\b}$; however, in both \cite{17} and 
\cite{18}-\cite{19} the boundary data imposed are not geometric, but incorporate or assume an extraneous 
choice of timelike unit vector $T$ along the boundary $\cT$. 

  To obtain a well-posed geometric IBVP, one expects that it is necessary to choose maximally dissipative boundary 
conditions on $\cT$, (cf.~also \cite{5} for discussion of various boundary conditions). The exact form of these 
will depend on the choice of gauge (and is currently unknown) but typically such boundary conditions 
have the schematic form
\be \label{17}
\del_{t}g_{ij} + \del_{\nu}g_{ij} = F_{ij},
\ee
where $\nu$ is the outward unit normal to $S \subset \S$ and $F_{ij}$ is given. Such boundary conditions are 
not close to Dirichlet (or Neumann) boundary conditions and so are far from the BY prescription. In fact, no 
component of the metric $\g$ itself appears in \eqref{17}. As discussed following \eqref{12}, the time-like 
unit normal $T$ to the Cauchy surface $\S$ is determined only globally, by solving the IBVP for given data 
\eqref{17}, and then reading off the value of $T$ at $S$. Only at that point can the Hamiltonian be actually 
computed as the unit-time variation of the action. (There is also the issue of finding a boundary term $B$ 
as in \eqref{15} so that boundary conditions analogous to \eqref{17} give a well-defined variational problem). 

\medskip

   The notion of quasi-local energy is difficult to make precise since energy is to be defined for an 
``isolated system" and it not a priori clear how to isolate a given region from its surroundings. 
(This issue bears some resemblance to certain versions of Mach's principle). Typically one would 
impose conditions such as no incoming radiation or absorbing boundary conditions, cf.~\cite{20} for example. 
However, due partly to the general covariance of GR, such boundary conditions are notoriously difficult 
to identify and implement in practice. Although simple and natural, it is unclear in what manner Dirichlet 
boundary conditions effectively model isolated physical systems. 

  There are now several very interesting and useful geometric notions associated to local spacetimes, such as 
the Hawking mass, the Brown-York energy, the Bel-Robinson energies and many others. These concepts are 
clearly very useful tools in understanding the physics, geometry and analysis of such spacetimes. However, 
as discussed above, they do not provide a fully satisfactory notion of quasi-local energy or Hamiltonian.

\medskip
 
   In closing, two brief remarks. First, the definition of the BY Hamiltonian, as well as its more recent 
modifications \cite{5}-\cite{8} require a choice of subtraction term to normalize the zero-point of the energy. These 
subtraction terms are typically determined by choices of isometric embedding of $(S, \g_{S})$ into either Euclidean 
space $\bR^{3}$ or Minkowski space $\bR^{1,3}$. On the other hand, one might hope that a correct choice of 
gauge and boundary conditions would obviate the need for such subtraction terms (which are somewhat 
artificial and adhoc given the intrinsic nature of GR). 

   Finally, the approach of Brown-York is used in the determination of the energy, mass and other charges for 
asymptotically AdS spacetimes in the AdS/CFT correspondence, cf.~\cite{21}, \cite{22} for instance. These 
concepts, which are of basic importance in the aspects of the correspondence related to thermodynamics of 
black holes, are global, and the charges are given by suitably renormalized integrals at conformal infinity. 
The difficulties discussed above in the quasi-local case do not apply in this context where limits at infinity 
are taken (as in the AF case). For example, in contrast to the finite case, the Dirichlet boundary value 
problem is well-posed at conformal infinity in the AdS context (for Euclidean metrics); the constraint 
equations "disappear" in the limit at infinity as restrictions on the form of the conformal metric at 
infinity.

\bibliographystyle{plain}

\end{document}